# Free energy landscape of two-state protein Acylphosphatase with large contact order revealed by force-dependent folding and unfolding dynamics

Xuening Ma[1#], Hao Sun [1,2,3#], Haiyan Hong[1], Zilong Guo[2,3], Huanhuan Su[1], Hu Chen[1,2,3]*

[1] Research Institute for Biomimetics and Soft Matter, Fujian Provincial Key Lab for Soft Functional Materials Research, Department of Physics, Xiamen University, Xiamen 361005, China
[2] Center of Biomedical Physics, Wenzhou Institute, University of Chinese Academy of Sciences, Wenzhou, China 325000
[3] Oujiang Laboratory, Wenzhou, Zhejiang 325000, China

\# Contributed equally to this work

\* E-mail: chenhu@xmu.edu.cn

**Abstract**

Acylphosphatase (AcP) is a small protein with 98 amino acid residues that catalyzes the hydrolysis of carboxyl-phosphate bonds. AcP is a typical two-state protein with slow folding rate due to its relatively large contact order in the native structure. The mechanical properties and unfolding behavior of AcP has been studied by atomic force microscope. But the folding and unfolding dynamics at low forces has not been reported. Here using stable magnetic tweezers, we measured the force-dependent folding rates within a force range from 1 pN to 3 pN, and unfolding rates from 15 pN to 40 pN. The obtained unfolding rates show different force sensitivities at forces below and above ~27 pN, which determines a free energy landscape with two energy barriers. Our results indicate that the free energy landscape of small globule proteins have general Bactrian camel shape, and large contact order of the native state produces a high barrier dominate at low forces.

# I. INTRODUCTION

The study of protein structure and function is the basis for understanding life phenomena at the molecular level. The protein obtains its functional structure through the folding process from random coils [1-3]. Misfolding of important proteins in cells can cause many diseases, such as Alzheimer's disease, mad cow disease, and Parkinson's disease [4-12].

Small single-domain proteins are the basic unit of proteins folding, as most multi-modular proteins are composed of these single-domain proteins. Conceptually the folding process of proteins is considered as a searching process on a funnel-shape free energy landscape [1, 13-15]. Most small single-domain proteins fold in a two-state manner, i.e., only the fully unfolded state and the native state are highly populated during the folding and unfolding process [16-18].

It has been found that the folding rates of small single-domain proteins are related to the topology of their native structures [19]. To grasp the basic topology of protein native structure, the average sequence separation between contacting residues in the native state is defined as contact order (CO). The proteins with small CO tend to fold faster than those with large CO. Therefore, it is fundamental to reveal the mechanism how CO affects the shape of free energy landscape, especially the location and height of transition barriers.

In traditional biochemical bulk experiments, denaturants like urea and Guanidine Hydrochloride (GmdCl) are introduced and removed to study the unfolding and refolding dynamics of proteins [20, 21]. The measured values reflecting the state population are always from the average results of large amount of protein molecules, from which the transition with transient intermediate state is difficult to study. Recently single-molecule experiments can detect the state of a single protein molecule, effectively overcoming the limitations of bulk experiments [22-24]. In single-molecule stretching experiments, the conformations of the protein and transitions are regulated by the tension applied to the end of the protein. For example, the folding rate and unfolding rate are both sensitive to the stretching force, and this force-dependence is related to the free energy landscape and transition pathways. Commonly used single-molecule manipulation techniques include atomic force microscope (AFM), optical tweezers and magnetic tweezers. Magnetic tweezers can exert an intrinsic constant pulling force on protein molecules, which does not require a complex feedback control system [25-27]. The stability of magnetic tweezers is significantly better than that of AFM and optical tweezers.

Acylphosphatase (AcP) is one of the smallest globular protein enzymes with about 98 amino acid residues, which catalyzes the hydrolysis of acylphosphates, and plays an important role in glycolysis, the tricarboxylic acid cycle, pyrimidine biosynthesis [28].

AcP is widely distributed in vertebrate species. It is found as two isoenzymes in many organs and tissues, called muscle type (mAcP) and organ common type (ctAcP). The structure of AcP consists of two α-helices packed against a five-stranded antiparallel β-sheet. Two β-α-β units are inserted into each other in the native state of AcP [29] [Fig. 1(a)]. AcP protein is also a model protein to study protein folding mechanism due to its small size, simple topology, and lack of intramolecular disulfide bonds. AcP has relatively large CO, and folds with folding rate of about 0.2 $s^{-1}$, much slower than other proteins with similar size [19], which makes AcP a good model protein to reveal the mechanism how CO affects the shape of free energy landscape.

In previous research, the folding transition state of mAcP was investigated by measuring the folding dynamics under various solvent conditions [30]. Heat- and urea-induced denaturation experiments investigated the thermodynamics and kinetics of folding of the ctAcP and the homologous mAcP. Both proteins fold in a two-state manner without intermediates [31-35]. Arad-Haase et al. did single molecular manipulation experiment by AFM to study the mechanical unfolding of polyprotein AcP under force [36]. They found that the unfolding rate at zero force is about 0.03 $s^{-1}$, which indicates that AcP is very unstable with folding free energy of only about 2-3 $k_BT$, which caused a contradiction with the stability of AcP [32]. Additionally, the force-dependent folding dynamics of AcP has never been studied.

Recently it has been found that the unfolding rate usually does not follow Bell's model in force range from several pN to tens or more than one hundred pN [17, 37, 38]. This is because that the rate-limiting free energy barrier is usually force-dependent. At smaller forces, the barrier with longer extension start to dominate the unfolding process, while at larger forces, the barrier with longer extension is suppressed and only the barrier with shorter extension plays as the main barrier of unfolding. As a slow folder with large CO, AcP is predicted to have a special free energy landscape which requires to be constructed based on the folding and unfolding dynamics in large force range.

In our work, taking advantage of the stability and force accuracy of magnetic tweezers, we measured the unfolding rate of the horse muscle AcP at 15-40 pN and the folding rate at 1-3 pN. The free energy landscape of AcP was constructed based on force-dependent folding and unfolding dynamics. Our results indicate that the large CO of AcP produces a higher free energy barrier with longer extension, which slows down the folding process of AcP.

# II. METHODS

## A. Sample preparation.

The gene of protein construct of HisTag-AviTag-FH1-AcP-FH1-SpyTag was synthesized and inserted into plasmid pET151 by GenScript. The above plasmid (ampicillin resistant) and another plasmid (chloramphenicol resistant) with gene of BirA (Biotin ligase) were mixed in an equal molar ratio and then transformed into the *E. coli* strain BL21. The transformed colonies were inoculated into LB medium (supplemented with chloramphenicol, ampicillin, D-biotin), and cultured at 37°C until the optical density (OD) of the bacterial cell reached 0.6. Then AcP protein was expressed for 12 h with 0.5 mM isopropyl-β-D-thiogalactopyranoside (IPTG) in medium at 25°C. The bacteria were collected by centrifugation at 4°C and lysed by sonication in a lysis buffer (50 mM Tris, 500 mM NaCl, 50% glycerol, 5 mM imidazole, 5 mM 2-mercaptoethanol, pH 8.0). The AcP protein was purified by using Ni-NTA Sefinose (TM) Resin (Sangon Biotech), and quick-frozen in liquid nitrogen and stored at -80°C [39].

## B. Single molecule measurement.

Coverslips of 22 mm × 32 mm and 22 mm × 22 mm were used to prepare the flow chamber. After cleaning by the sonicator and oxygen plasma cleaner, the glass slides were silanized by immersing in a methanol solution of 1% 3-aminopropyltriethoxysilane (APTES, cat. A3648, Sigma) for 1 h and rinsed by methanol. Flow chambers were formed by sticking parafilm between a piece of APTES-functionalized coverslip and another piece of cleaned coverslip. Polybead Amino Microspheres (cat. 17145, Poly-sciences) with a diameter of 3.0 μm were flowed into the chamber and incubated for 20 minutes to stick on the surface of the coverslip and used as a reference to eliminate the spatial drift during the experiment. After rinsing with 200 μL PBS buffer, 1% Sulfo-SMCC (SE 247420, Thermo Science) was flowed into the chamber and incubated for 20 minutes. After that, N-SpyCatcher protein in PBS was flowed into chamber and incubated for 1 h. Then 1% BSA in 1×PBS was flowed into the chamber and incubated overnight to passivate the surface. The appropriate concentration of protein with N-terminal AviTag and C-terminal SpyTag in 1% BSA solution in 1×PBS was flowed into the chamber and incubated for 20 minutes. The streptavidin-coated paramagnetic beads Dynabead M270 were flowed into the chamber to form protein tethers. Sketch of the protein tether is shown in Fig. 1(a). Finally, 1×PBS buffer with 1% BSA and 10 mM sodium ascorbate was flowed into the chamber.

The magnetic tweezers were built on an inverted microscope. 100X oil immersion objective was used to take images of protein-tethered bead. The force-dependent folding and unfolding dynamics of AcP were measured through force-jump experiments [38] which can be achieved by moving the magnets rapidly in less than 0.15 s. The design details of magnetic tweezers can refer to our previous publications [26].

# III. RESULTS

## A. Force extension curve at constant loading rate identifies the correct unfolding signal of AcP.

In magnetic tweezers stretching experiments, AcP with FH1 linkers and biotin at the N-terminus and SpyTag at the C-terminus was successfully attached to streptavidin-coated paramagnetic bead and SpyCatcher-coated coverslip [40] [Fig. 1(a)]. Stretching force increased with a constant loading rate of 1 pN/s from 0 pN to 30 pN. Two types of typical signals were recorded in the force extension curve. Firstly, a distinct 22 nm unfolding step was observed under ~19 pN, which gives a contour length increase of about 36 nm based on worm-like chain (WLC) estimation [Fig. 4(c)]. This unfolding step size is consistent with the unfolding of the target protein AcP. When force was increased to about 30 pN, several equilibrium folding and unfolding steps with size of ~3.5 nm were recorded, which is the fingerprint signal from SpyTag/SpyCatcher complex [39] [Fig. 1(b)]. Hereafter, all protein tethers were verified by this typical 3.5 nm fingerprint signal before subsequent force-jump measurements.

## B. Force-jump experiment to explore the unfolding rates of AcP.

The force-dependent unfolding rates of AcP was explored by force-jump measurement in the force range from 15 pN to 40 pN. Multiple cycles of different force jump experiments were done to collect enough statistical experimental results. In each force cycle, a low force of 0.5 pN was applied for 20 s to make the protein successfully fold to its native state. Then force was increased abruptly to a big value from 15 pN to 40 pN, and hold for certain duration for the protein to unfold, which is 60 s for 15 pN and 20 pN, 25 s for 25 pN, 7 s for 30 pN, 5 s for 35 pN and 40 pN. At forces of 15 pN and 20 pN, the unfolding of AcP was not observed every time in the duration of 60 s. In this case, the probability of unfolding at 15 pN in 60 s was 14%, and that at 20 pN in 60 s was 76% [Fig. 2(a)].

As shown in the Fig. 2(b), the survival probability of native state of AcP at 20 pN and 35 pN as a function of the unfolding time was obtained from cumulative distribution of the lifetime of the native state, the unfolding rate $k_u$ was determined by exponential fitting. The survival probability data to obtain $k_u$ of AcP under other forces were given in Fig. S1.

## C. Force-jump experiment to explore the folding rates of AcP.

We find that the unfolded AcP can refold to its native state when force is below 3 pN, but we cannot identify a visible folding step during the experiment, as the large extension fluctuation at low forces obscures the folding signal. We designed a special experimental procedure to measure the folding rate at forces below 3 pN [41]. First, force of 30 pN was applied for 7 s to unfold AcP protein, then force was dropped to 1 pN and hold for 2 s for AcP to refold, after that force of 30 pN was applied again to check if unfolded AcP refolds successfully from the extension at 30 pN. Before each measurement, the folding capability of AcP was checked by applying lower force of 0.5 pN for 20 s [Fig. 3(a)]. Different folding time at 1 pN was used in the force cycle, and the folding probability of different folding time was calculated and shown in Fig. 3(b). Exponential fitting gives the folding rate $k_f$ of AcP at 1 pN. Similarly, the folding rates of AcP at 2 pN and 3 pN were also obtained by the above method [Fig. 4(b), Fig. S2].

## D. Theoretical analysis of the force-dependent folding and unfolding rates.

In Fig. 4(a), the force-dependent unfolding rate and folding rate obtained from force-jump measurement were plotted with logarithm scale. Data points for unfolding rate were obtained from six independently measured tethers [Fig. S3]. The force-dependent unfolding rates are not along a straight line. Therefore, they cannot be described by Bell's model [42]: $k_u(f) = k_u^0 \exp(f x_u / k_B T)$. Here, $f$ is the tensile force, $k_u^0$ is the unfolding rate at zero force, $x_u$ is the unfolding transition distance from the native state to the transition state, $k_B$ is the Boltzmann constant, and T is the absolute temperature.

Noteworthy, the unfolding rates can be well fitted by two Bell's models at 15-25 pN and 30-40 pN, respectively. The unfolding rates from 30 pN to 40 pN can be fitted with Bell's model with fitting parameters $k_{u,1}^0 = (6 \pm 3) \times 10^{-3}$ s$^{-1}$ and $x_{u,1} = 0.69 \pm 0.06$ nm, which is consistent to the result from previous AFM experiment [36]. Fitting with Bell's model from 15 pN to 25 pN gives $k_{u,2}^0 = (2.8 \pm 0.6) \times 10^{-6}$ s$^{-1}$ and $x_{u,2} = 1.85 \pm 0.06$ nm. The intersection of two fitted lines with different slopes was located at about 27 pN [Fig. 4(a)].

As there are two barriers with $x_{u,1} = 0.69$ nm and $x_{u,2} = 1.85$ nm, we suppose that there is a transient intermediate state I between these two barriers. Therefore, the two state model need to be revised to include I state [17]: $N \rightleftharpoons I \to U$ to describe the unfolding process. The measured unfolding rates in full experimental force range 15-40 pN can be fitted with the analytical equation of the NIU model [43] [dashed line in Fig. 4(a)]. Interdependence between parameters about the intermediate state were described in Fig. S4.

The folding rate decreases drastically as force increases. Over a small force range from 1 to 3 pN, $k_f(f)$ decreases from ~ 0.1 to ~ 0.009 s$^{-1}$. The force-dependent folding rate can be fitted by equation of Arrhenius' law: $k_f(f) = k_f^0 \exp\left(-\int_0^f x_f(f') df' / k_B T\right)$, where $k_f^0$ denotes the folding rate at zero force, $x(f) = x_{\text{chain}}(f) - x_{\text{TS}}(f)$ is the extension difference between the unfolded state ($x_{\text{chain}}(f)$) and the folding transition state ($x_{\text{TS}}(f)$). $x_{\text{chain}}(f)$ is determined by WLC force-extension equation: $\frac{fA}{k_B T} = \frac{x_{\text{chain}}}{L} + \frac{1}{4}\left(1 - \frac{x_{\text{chain}}}{L}\right)^{-2} - \frac{1}{4}$, where persistence length of A ~ 0.8 nm at low forces and contour length of L ~35.77 nm (0.365 nm per amino acid and 98 amino acids). We suppose that the folding transition state can be approximated modeled as a rigid body. Under the force, the tethered rigid body can only rotate to align with the force direction with thermal fluctuations. Its extension can be described as the same analytical solution as the monomer force-extension curve in the free-joint chain (FJC) model [26]: $x_{\text{AcP}}(f) = l_0 \coth\left(\frac{fl_0}{k_B T}\right) - \frac{k_B T}{f}$, where $l_0$ is the N-C distance of this folding transition state. A zero-force folding rate $k_f^0 = 0.14$ s$^{-1}$ and the size of folding transition state $l_0 = 5.0$ nm can be obtained by fitting the experimental data [Fig. 4(b)].

The unfolding step size under different forces obtained in the force-jump experiments was shown in Fig. 4(c), which is consistent with the theoretical difference between $x_{\text{chain}}(f)$ and $x_{\text{AcP}}(f)$. Within the error range, the coincidence of experimental values and theoretical values further proves the accuracy of our experimental data.

### E. Free energy landscape construction.

The end-to-end distance of AcP on the basis of the crystal structure of AcP is ~2.5 nm (PDB:1APS). Two different unfolding distances ($x_{u,1} = 0.69$ and $x_{u,2} = 1.85$ nm) were determined by fitting the unfolding rates in two different force ranges [Fig. 4(a)], which means that there are two transition states (TS) at N-C distance of 3.2 nm (TS1) and 4.4 nm (TS2) [Fig. 5]. Based on the free-joint chain model of a peptide, the size of the unfolded state can be estimated to be ~7.5 nm. We inferred that there is an intermediate state between TS1 and TS2, which is actually not observed in the experimental extension time course.

Based on the force-dependent folding rates and unfolding rates, we can determine the unfolding free energy barrier at zero force according to an empirical equation: $k_u^0 = k^* \exp(-\Delta G^\ddagger / k_B T)$, where we suppose that the intrinsic transition rate $k^*$ is $10^6$ s$^{-1}$, then the two free energy barriers are calculated as ~18.9 $k_B$T and ~26.6 $k_B$T for TS1 and TS2, respectively.

From the folding and unfolding measurements, the critical force $f_c$ at which the protein has equal folding and unfolding rates is between 3 pN and 15 pN. But we cannot get a reliable equilibrium measurement because the transition rates are very slow. Based on the extrapolation from force-dependent folding rates and unfolding rates, the critical force is estimated to be ~5.3 pN [Fig. 4(a)]. The folding free energy of protein $\Delta G_0 = \int_0^{f_c} \Delta x(f') df' = 10.9 \, k_B T$, where $\Delta x(f) = x_{\text{chain}}(f) - x_{\text{AcP}}(f)$ is the force-dependent unfolding step size. This folding free energy is consistent with previous bulk experiment [32].

## IV. DISCUSSION

AcP folds slowly due to its large CO in the native state. From the force-dependent folding rates at forces 1-3 pN, the zero-force folding rate is obtained as 0.14 s$^{-1}$, which is consistent with previous bulk measurement [30, 32]. This slow folding rate indicates that AcP must overcome a high folding barrier from the unfolded state to its native state. It costs a significant entropy penalty for two residues far away from each other in amino acids sequence to meet, which produces a higher free energy barrier.

In AFM experiment, as the unfolding force is bigger than 25 pN, the resulting unfolding distance is only 0.6 nm, thus it is hardly to make sure that this unfolding barrier is the same as the folding barrier at zero force [36]. In this work we measured the unfolding rates from 15 pN to 40 pN. Force-dependent unfolding rates at forces greater than 27 pN is consistent with previous AFM experiment, while the unfolding rate becomes much more sensitive to force when force is smaller than 27 pN. In the plot of logarithm of unfolding rate vs. force, the slope in the force range below 27 pN is much larger than the slope at forces greater than 27 pN. Therefore, these two distinct slopes indicate that there are two barriers along the unfolding pathway. At large forces, the barrier with longer extension TS2 is suppressed, and the rate-limiting barrier is the one with small extension TS1, whose conformation is very close to the native state. While at low forces or zero force, TS2 becomes the barrier with the highest free energy, which makes it the main barrier for protein to cross during the unfolding process.

Protein folding and unfolding are reversible processes. From our force-dependent folding rate measurement in the force range from 1 to 3 pN, the size of the folding transition state is estimated to be about 5 nm. The extension of this folding transition state is similar to the extension of the unfolding transition state measured in the force range from 15 to 25 pN. Therefore, the main unfolding barrier TS2 at low forces is the candidate of the folding barrier.

In general, the founding for AcP agrees with the force-dependent unfolding and folding dynamics of several other two-state proteins like GB1, CSP, and I27, which all show deviation from Bell's model when the force range expands to low force regime from the typical AFM force range of tens of pN [17, 26, 38]. AcP has large CO, and consequently its dominant barrier at small force has higher free energy. We suppose that the remote contacting residues in AcP start to move close to each other when crossing the barrier TS2, and the large entropy cost to form large loops contributes to the high free energy of TS2. If the hidden intermediate state between TS1 and TS2 is the theoretically proposed molten globular state, then it must be a loosely compact conformation with topology similar to the protein's native state.

# ACKNOWLEDGEMENTS

This work was supported by the National Natural Science Foundation of China (Grant No. 11874309 and 12174322) and 111 project (B16029).

# FIGURES

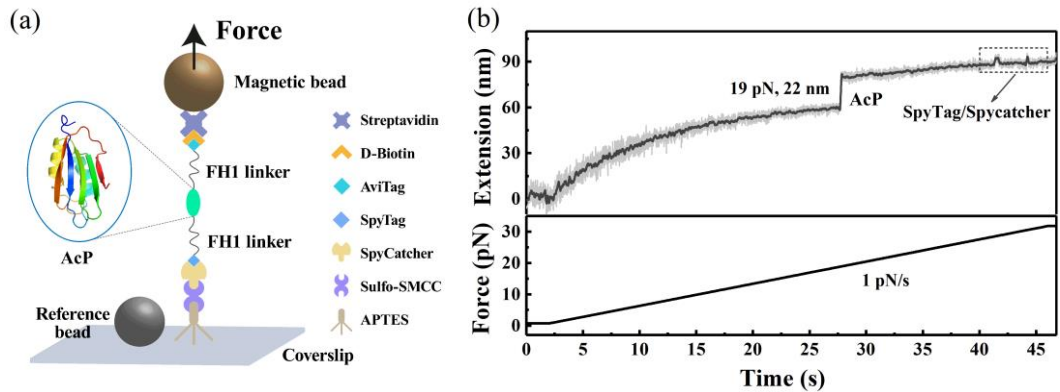

FIG. 1. Sketch of magnetic tweezers measuring the force response of AcP protein and representative experimental data. (a) Schematic of protein-tethered magnetic bead: Construct of AviTag(biotin)-FH1-AcP-FH1-SpyTag was attached to streptavidin-coated paramagnetic bead and the SpyCatcher-coated coverslip surface. Zoomed-in shows the crystal structure of AcP (PDB:1APS). (b) Force-extension curve shows AcP unfolding with step size of 22 nm at force of ~19 pN during the force-increasing process with a loading rate of 1 pN/s. The dashed frame shows the unfolding-folding transitions of SpyTag/SpyCatcher. Raw data were recorded at 200 Hz (grey line) and smoothed in a time window of 0.25 s (black line).

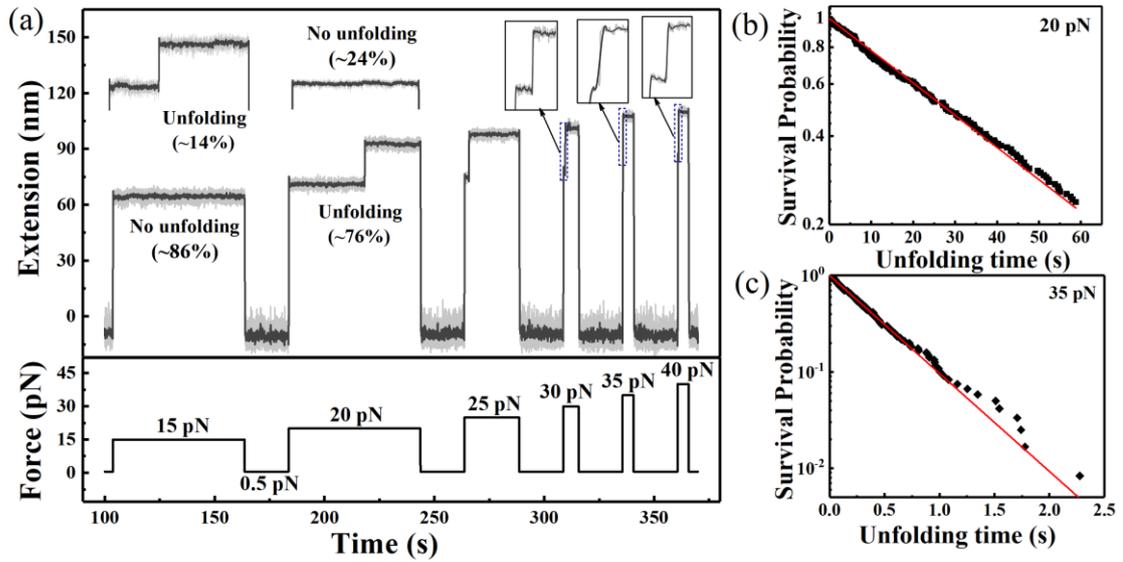

FIG. 2. Unfolding dynamics of AcP under different forces in the force-jump experiment. (a) Typical force-jump measurement of the unfolding process of AcP. It shows an example of six force-jump cycles on the same protein tether. Before each cycle, the protein folded to its native state at a force of 0.5 pN for 20 s. When applying forces of 15 or 20 pN for 60 s, we can see the unfolding step of AcP with probabilities of ~14% and ~76%, respectively. When applying forces of 25 pN for 25 s, and 30 pN for 7 s, 35 pN for 5 s, 40 pN for 5 s, unfolding transitions were recorded in all force cycles. Raw data were recorded at 200 Hz (grey line) and smoothed in a time window of 0.25 s (black line). (b) The survival probability of native state of AcP at 20 pN and 35 pN. The red solid line shows exponential fitting curve to determine $k_u$ of AcP. Similarly, $k_u$ of AcP under other forces was determined [Fig. S1].

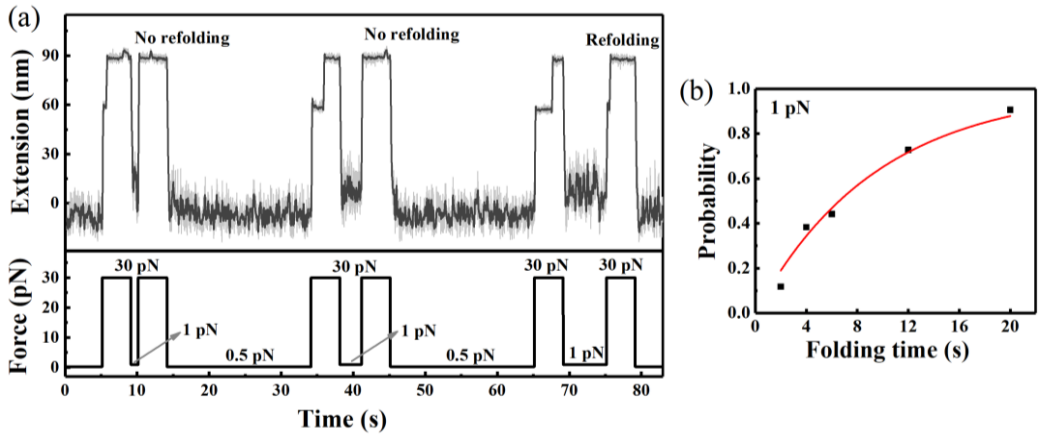

FIG. 3. Measurement of AcP folding rate. (a) The measurement of the folding rate at 1 pN. Firstly, force of 30 pN was applied to unfold AcP. Then force was dropped to 1 pN and let protein to fold at 1 pN for 2 s. Due to large fluctuation of extension at 1 pN, we cannot observe the clear folding process directly. Therefore, we increased force to 30 pN and recorded the extension time course from which we determine if the folding at 1 pN for 2 s was successful or not. The same measurement was done for 4 s, 6 s, 12 s, and 20 s. Before each measurement, 0.5 pN was applied for 20 s to make sure the protein can successfully fold. Raw data were recorded at 200 Hz (grey line) and smoothed in a time window of 0.25 s (black line). (b) The folding probability at different folding time at 1 pN. Exponential fitting determines the folding rate of AcP as 0.11 s$^{-1}$ at 1 pN (red solid line).

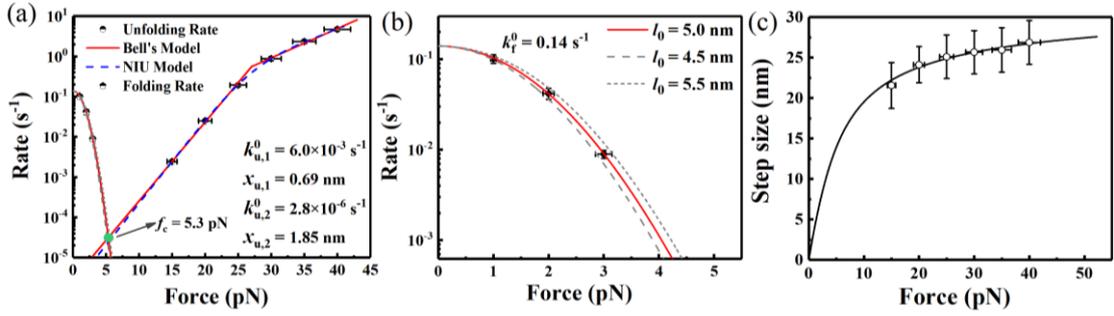

FIG. 4. Force-dependent folding/unfolding rates and unfolding step sizes of AcP. (a) The folding rates and unfolding rates of AcP were obtained from force-jump experiment. The average unfolding rates were fitted with Bell's model in the force ranges from 15 to 25 pN and from 30 to 40 pN, which determined two different $x_u$: $x_{u,1} = 0.69 \pm 0.06$ nm, $x_{u,2} = 1.85 \pm 0.06$ nm, and corresponding parameters $k^0_{u,1} = (6 \pm 3) \times 10^{-3}$ s$^{-1}$, $x_{u,1} = 0.69 \pm 0.06$ nm, respectively. Critical force of 5.3 pN is estimated from the extrapolation of force-dependent folding/unfolding rates. The unfolding rate came from six different tethers and the folding rate came from three different tethers [Fig. S3]. Error bar is obtained from the standard error of the mean lifetime. Force is estimated to have 5% uncertainty. (b) Force-dependent folding rates are fitted by Arrhenius' law to determine the size of folding transition state as $5.0 \pm 0.5$ nm. (c) The average unfolding step sizes of AcP were obtained from force-jump measurement. Error bar is the standard deviation. The black curve represents the worm-like chain fitting with contour length of ~35.8 nm and persistence length of ~0.8 nm.

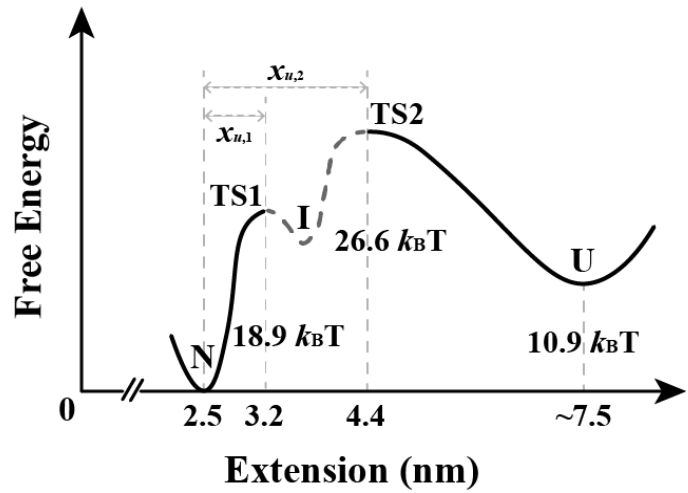

FIG. 5. Free energy landscape of AcP constructed from force-dependent folding and unfolding rates. Folding free energy of AcP of 10.9 $k_B T$ was obtained from the extrapolated critical force of 5.3 pN [Fig. 4(a)]. Unfolding distances $x_{u,1}$ and $x_{u,2}$ were obtained from the slope of force-dependent unfolding rate at forces greater than 30 pN and smaller than 25 pN, respectively. Unfolding barriers were estimated based on an assumption of the intrinsic unfolding rate of $10^6$ $s^{-1}$. The dashed line between TS1 and TS2 represents an unstable intermediate state which was not observed in the extension time course. Size of the unfolded state is estimated from the room mean square of the end-to-end distance of random coil polypeptide.

# Supplementary Information for

# "Free energy landscape of two-state protein Acylphosphatase with large contact order revealed by force-dependent folding and unfolding dynamics"

Xuening Ma[1#], Hao Sun [1,2,3#], Haiyan Hong[1], Zilong Guo[2,3], Huanhuan Su[1], Hu Chen[1,2,3]*

[1] Research Institute for Biomimetics and Soft Matter, Fujian Provincial Key Lab for Soft Functional Materials Research, Department of Physics, Xiamen University, Xiamen 361005, China

[2]Center of Biomedical Physics, Wenzhou Institute, University of Chinese Academy of Sciences, Wenzhou, China 325000

[3]Oujiang Laboratory, Wenzhou, Zhejiang 325000, China

\# Contributed equally to this work

\* E-mail: chenhu@xmu.edu.cn

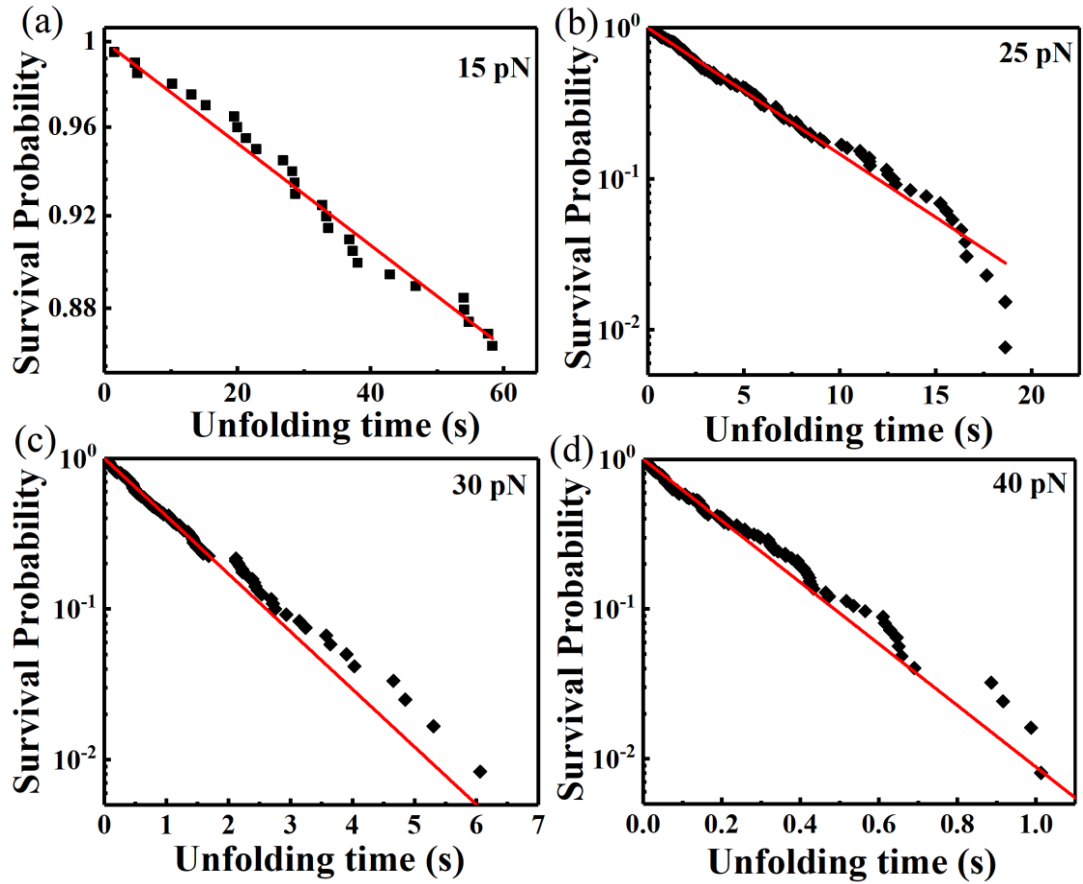

FIG. S1. The survival probability of native state of AcP at 15 pN, 25 pN, 30 pN and 40 pN. The red solid line shows exponential fitting curve to determine the unfolding rate of AcP. Statistics of data points come from six independently-measured tethers.

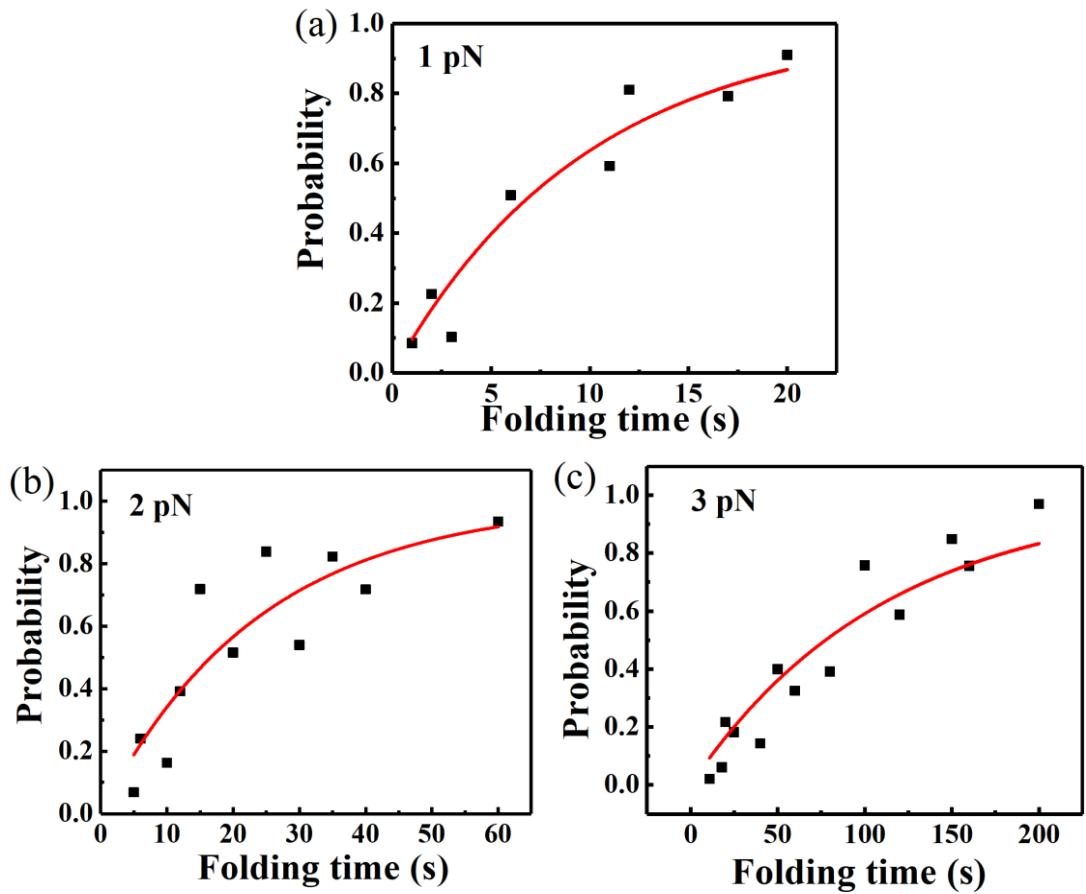

FIG. S2. Data points from three independently-measured tethers represent folding probability at different folding time, and the exponential fitting curves determine the folding rates of AcP as 0.10 $s^{-1}$ at 1 pN, 0.04 $s^{-1}$ at 2 pN, 0.009 $s^{-1}$ at 3 pN (red solid line).

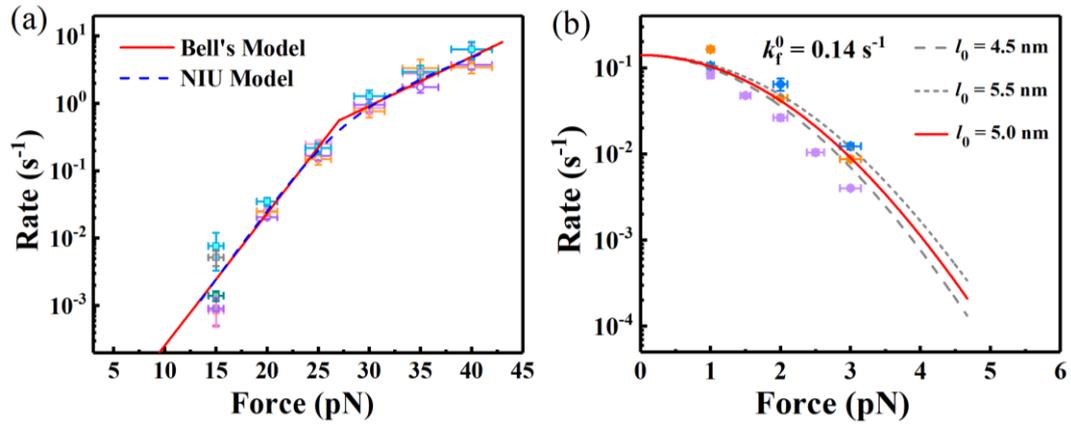

FIG. S3. Force-dependent unfolding and folding rates from individual protein tether. (a) Force-dependent unfolding rates from six independently-measured protein tethers are plotted with different symbols. Circular data points are determined by a single exponential fitting to the survival probability of native state at different forces. Square data points are the reciprocal of the mean unfolding time. (b) Force-dependent folding rates from three independently-measured tethers. Fitting curves in Fig. 4 are shown in both (a) and (b).

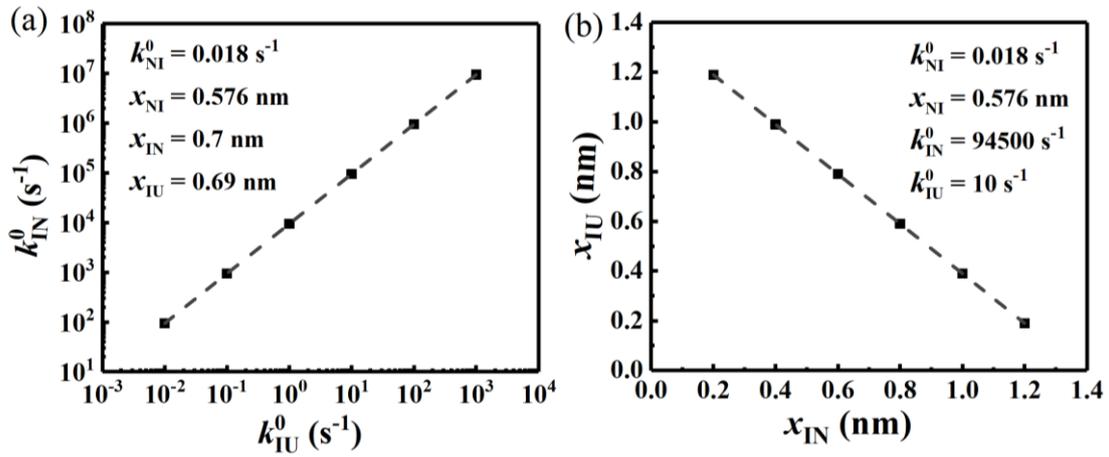

FIG. S4. Using the NIU model to fit the force-dependent unfolding rates, we found the interdependence among the parameters. Detailed equation can be found in reference [1]. (a) Fixing the parameter values in the figure, parameters $x_{IU}$ and $x_{IN}$ are anti-correlated to each other. (b) Fixing the parameter values in the figure, parameters $k_{IN}^0$ and $k_{IU}^0$ are proportional to each other.